# Existence of the upper critical dimension of the Kardar-Parisi-Zhang equation

Eytan Katzav and Moshe Schwartz

Raymond and Beverly Sackler Faculty of Exact Sciences

School of Physics and Astronomy, Tel Aviv University,

Ramat Aviv, Tel Aviv 69978, Israel

Abstract

The controversy whether or not the Kardar-Parisi-Zhang (KPZ) equation has an upper critical dimension (UCD) is going on for quite a long time. Some approximate integral equations for the two-point function served as an indication for the existence of a UCD, by obtaining a dimension, above which the equation does not have a strong coupling solution. A surprising aspect of these studies, however, is that various authors that considered the same equation produced large variations in the UCD. This caused some doubts concerning the existence of a UCD. Here we revisit these calculations, describe the reason for such large variations in the results of identical calculations, show by a large-$d$ asymptotic expansion that indeed there exist a UCD and then obtain it numerically by properly defining the integrals involved. Since many difficult problems in condensed matter physics of non-linear nature are handled with mode-coupling and self-consistent theories, this work might also contribute to other researchers working on a large class of different problems that might run into the same inconsistencies.

The KPZ equation [1] for surface growth under ballistic deposition was introduced as an extension of the Edwards-Wilkinson theory [2]. The interest in the KPZ equation exceeds far beyond the interest in evolving surfaces because of the following reasons: (a) The KPZ system is known to be equivalent to a number of very different physical systems. Examples are the directed polymer in a random medium and the Schrodinger equation (in imaginary time) for a particle in the presence of a potential that is random in space and time. (b) The second reason, that is more important, to our mind, is that it serves as a relatively simple prototype of non-linear stochastic field equations, that are so abundant in condensed matter physics.

The equation for the height of the surface at the point $\vec{r}$ and time t, $h(\vec{r},t)$, is given by

$$\frac{\partial h}{\partial t} = \nu \nabla^2 h + \frac{\lambda}{2}(\nabla h)^2 + \eta(\vec{r},t), \qquad (1)$$

where $\eta(\vec{r},t)$ is a noise term such that

$$\langle \eta(\vec{r},t) \rangle = 0 \quad \text{and} \quad \langle \eta(\vec{r},t)\eta(\vec{r}',t) \rangle = 2D\delta(\vec{r}-\vec{r}')\delta(t-t'), \qquad (2)$$

so that the constant deposition rate is removed.

One of the quantities of interest is the roughness exponent $\alpha$, that characterizes the surface in steady state. It is defined by

$$\left\langle \left[h(\vec{r}) - h(\vec{r}')\right]^2 \right\rangle \propto |\vec{r}-\vec{r}'|^{2\alpha}. \qquad (3)$$

It is well known that for dimensions $d \leq 2$ the surface is always rough ($0 < \alpha < 1$), and the two point function $\langle h_q h_{-q} \rangle$ ($h_q$ being the Fourier transform of $h(\vec{r})$) cannot be obtained perturbatively. Above two dimensions, that is known as the lower critical dimension (LCD), the picture changes and there is a transition depending on the physical parameters, from a non-perturbative strong coupling regime, that is rough, to a weak coupling regime, characterized by a flat surface, that can be obtained perturbatively, and its leading behavior is given by the Edwards-Wilkinson model.

There is however a long lasting controversy regarding the existence of an upper critical dimension (UCD) above which the surface is always flat [3]. Various approaches, applied either to the KPZ equation directly or to the equivalent directed polymer problem, suggest that a UCD does not exist. An important indication for that comes from numerical simulations [4] which yields $\alpha > 0$ at least up to $d = 7$, well above any value obtained for the UCD in the literature. This result is also corroborated by two real-space renormalization group approaches [5, 6], that predict no UCD as well. On the other hand, finite UCD's are obtained by various field theoretical treatments [7-13], and the $1/d$ expansion of Cook and Derrida [14]. The first to obtain values for the UCD from approaches that gained credibility, by being able to give good values for $\alpha$ for d = 2, were Bouchaud and Cates (BC) [11]. They considered the equation obtained by Schwartz and Edwards (SE) [8] for the roughness exponent and obtained a UCD above which there is no strong-coupling solution. The value they give for the UCD arising from the SE method is 2.78. The value they obtain for the UCD from their own mode-coupling equation is 2.85 (in an erratum published a few months later, they replaced these UCD's with 3.25 and 3.75 respectively [15]). A more recent result worthwhile mentioning is that of Lassig & Kinzelbach [3] that predicted $d_c = 4$, and in addition were able to prove on a quite general assumption that anyway $d_c \leq 4$.

Motivation for the present paper lies in the article of Blum and McKane [10] that recalculated the upper critical dimension and obtained using the SE method that the UCD is 3.2 and using the BC method that the UCD is 3.6. Being unaware at the time of the erratum of BC [15], we believed that such discrepancies, cannot be attributed to slight numerical inaccuracies, and that led us to some numerical experimentation with both equations, and we were surprised to discover that indeed the variations in UCD's obtained by using identical methods may be large. This suggested at once three questions: (a) What is the reason for these discrepancies? (b) Can the results of a finite UCD be trusted at all? And If the answer to the last question is positive, (c) what are the correct values predicted by those methods? It turns out that the above questions and the conclusions that follow are relevant even in view of the erratum of BC. In fact this discussion extends well beyond the KPZ equation to the growing class of mode-coupling and self-consistent theories, applied to many non-linear problems. In addition to answering the above questions we had in mind the need to establish rigorous foundations for these theories in order not to face again these misfortunate inconsistencies, thus to enable easier development for them.

Our conclusions to be described later in detail are the following:

(a) One possible reason for the discrepancies is that both conclusions (BC and SE) involve the evaluation of two-dimensional integrals that are not absolutely convergent. Therefore, different ways of evaluating the integrals yield different results.

(b) This can be corrected by introducing a high momentum cutoff in the original KPZ equation, and letting eventually this cutoff to tend to infinity. This defines a unique way of evaluating the integrals that may be described as physically correct.

(c) Even if the correct integration scheme is used, there is still an interesting difficulty. The upper critical dimension should be the dimension where the roughness exponent, $\alpha$, becomes zero. It turns out, however, that the integrals under consideration are not continuous as a function of α. If we denote such an integral by I(d, α) we find that

$$\lim_{\alpha \to 0+} I(d,\alpha) \neq I(d,0). \quad (4)$$

This immediately implies that in the equation that should determine the upper critical dimension (which is an equation for d) α is <u>not</u> to be set to zero but the limit α → 0+ has to be taken.

(d) Taking into account the previous remarks we find that indeed the SE and BC method yield upper critical dimensions that are 3.7 and 4.3 respectively. This is certainly a big difference in the predicted $d_c$'s. In fact, it is so big that while in refs. [10] and [15] the BC results obey the Lassig inequality $d_c \leq 4$ [3], the proper evaluation of the integrals in our paper shows that the critical dimension predicted by the BC equations violates this inequality. Thus, the proper evaluation of the integrals shows that the BC equations are not consistent with Lassig's result, while the SE result is.

In the following we detail the reasons for our conclusions.

In the SE method [8-9], the roughness exponent $\alpha$ is determined by solving the transcendental equation $F(d,\alpha) = 0$, where

$$F(d,\alpha) \equiv -\int d^d t \frac{\vec{t}\cdot(\hat{e}-\vec{t})}{\left[t^{2-\alpha} + \left|\hat{e}-\vec{t}\right|^{2-\alpha} + 1\right]} \left[\vec{t}\cdot\hat{e}\cdot t^{-(d+2\alpha)} + (\hat{e}-\vec{t})\cdot\hat{e}\cdot\left|\hat{e}-\vec{t}\right|^{-(d+2\alpha)}\right] +$$
$$+ \int d^d t \frac{\left[\vec{t}\cdot(\hat{e}-\vec{t})\right]^2}{\left[t^{2-\alpha} + \left|\hat{e}-\vec{t}\right|^{2-\alpha} + 1\right]} \cdot t^{-(d+2\alpha)} \cdot \left|\hat{e}-\vec{t}\right|^{-(d+2\alpha)}, \quad (5)$$

where $\hat{e}$ is a unit vector in an arbitrary direction and the $\vec{t}$ integration is over all d-dimensional space.

The BC equation [11] is similarly $G(d,\alpha) = 0$, where in the definition of G the denominator $\left[ t^{2-\alpha} + |\hat{e} - \vec{t}|^{2-\alpha} + 1 \right]$, appearing in the definition of F, is replaced by $\left[ t^{2-\alpha} + |\hat{e} - \vec{t}|^{2-\alpha} \right]$.

From the definition of $\alpha$ it is clear that $\alpha$ must be less than 1. It turns out that for such $\alpha$'s the first two integrals in eq. (5) (defined by the two terms in the square brackets in the numerator) do not converge absolutely. It is known that an integral, which does converge absolutely, can be summed up in different ways, resulting in different values. This ambiguity calls for a clear and natural definition of how exactly this integral should be performed.

A physically natural way of defining the integrals is to assume that the original KPZ equation has a high momentum cutoff corresponding to a finite size of the grains poured onto the surface. The integrals have to be performed under this assumption and the size of the grain is then taken to zero.

The effect of a high momentum cutoff is to multiply the integrand in the integrals on the right hand side of eq. (5) by $\Theta(R - t) \cdot \Theta(R - |\hat{e} - \vec{t}|)$, where R is the parameter that has to be taken eventually to infinity (clearly a soft rather than an abrupt cutoff leads to the same results). It is amusing to consider the two contributions to the first integral on the right hand side of eq. (5). These should be identical, because one is transformed into the other by the transformation $\vec{t} \leftrightarrow \hat{e} - \vec{t}$. The actual values of the two contributions depend on the way the integrals are performed. If we introduce the cutoff as described above the two contributions are identical, but if we do the angular integration first and then do the t-integration from 0 to R, letting eventually R tend to infinity (In fact, it can be shown that in calculating $I_1$ or $J_1$ we use only $\Theta(R-t)$ as a cutoff factor in both integrals. It can be shown that in these cases it is enough to use only the factor $\Theta(R-t)$ as a cutoff), we obtain different results for the two contributions. This is not to say that the natural cutoff procedure is the only one that yields identical results for those two contributions.

In order to continue our discussion, let us now apply the above definition for the integration procedure, and use a more convenient form for the function $F(d,\alpha)$ of the SE equation [8, 9] by defining

$$I_1(d,\alpha) \equiv \int d^d t \frac{[\vec{t} \cdot (\hat{e}-\vec{t})][\vec{t} \cdot \hat{e}] \cdot t^{-(d+2\alpha)}}{\left[t^{2-\alpha} + |\hat{e}-\vec{t}|^{2-\alpha} + 1\right]}, \qquad (6)$$

and

$$I_2(d,\alpha) \equiv \int d^d t \frac{[\vec{t} \cdot (\hat{e}-\vec{t})]^2 \cdot t^{-(d+2\alpha)} \cdot |\hat{e}-\vec{t}|^{-(d+2\alpha)}}{\left[t^{2-\alpha} + |\hat{e}-\vec{t}|^{2-\alpha} + 1\right]}, \qquad (7)$$

we arrive at the equivalent SE equation

$$F(d,\alpha) = -2I_1(d,\alpha) + I_2(d,\alpha) = 0. \qquad (8)$$

Similarly, by defining $J_i(d, \alpha)$ (i = 1,2) the same as $I_i(d, \alpha)$, except that the term in the denominator $\left[t^{2-\alpha} + |\hat{e}-\vec{t}|^{2-\alpha} + 1\right]$ is replaced by $\left[t^{2-\alpha} + |\hat{e}-\vec{t}|^{2-\alpha}\right]$, we can reformulate the BC equation [11] as

$$G(d,\alpha) = -2J_1(d,\alpha) + J_2(d,\alpha) = 0, \qquad (9)$$

and the four additional schemes for determining $\alpha$ in the article of Blum and McKane [10] as

(scheme 1) $\qquad \partial_\alpha \left[-2I_1(d,\alpha) + I_2(d,\alpha)\right] = 0, \qquad (10)$

(scheme 2) $\qquad \partial_\alpha \left[-4J_1(d,\alpha) + J_2(d,\alpha)\right] = 0, \qquad (11)$

(scheme 3) $\quad \partial_\alpha [I_2(d,\alpha)] = 0$, (12)

(scheme 4) $\quad \partial_\alpha [J_2(d,\alpha)] = 0$. (13)

where $\partial_\alpha$ means differentiating with respect to $\alpha$.

Suppose next that the SE equation does not yield an upper critical dimension. In such a case we would expect a roughness exponent $\alpha$ that vanishes as the dimension d tends to infinity. (This expectation is consistent with numerical simulations as well as with almost all the approaches predicting no upper critical dimension, that usually predict $\alpha \sim 1/d$.) Consequently, the SE equation (eq. 8) for large d and small $\alpha$ must take the form

$$f_1(d,\alpha) - f_2(d) = 0, \quad (14)$$

where for $\alpha > 0$ $\lim_{d \to \infty} \dfrac{f_2(d)}{f_1(d,\alpha)} = 0$ and $f_1(d,\alpha)$ vanishes when $\alpha$ is set to be zero.

Therefore, we derive in appendix A the large d asymptotic expansion for the integrals $I_1(d,\alpha)$, $I_2(d,\alpha)$, $J_1(d,\alpha)$ and $J_2(d,\alpha)$.

In appendix A we also obtain the leading order dependence for $I_1(d,0)$ and $I_2(d,0)$. We find that $f_1(d,0) = -2I_1(d,0) + I_2(d,0) = -\dfrac{\Omega(d-1)}{4} \dfrac{\sqrt{2\pi}}{d^{3/2}} \neq 0$. It is thus clear that the condition for having a small $\alpha$ solution for large d is not obeyed. The case of the BC equation is a bit more interesting, because in this case we find that the leading order contribution $-2J_1(d,\alpha) + J_2(d,\alpha)$ vanishes when $\alpha = 0$. In order to see whether a small $\alpha$ solution exists for large d, a more general examination is needed. The argument following eq. (14) assumes that $f_1(d,\alpha)$ is continuous in the limit $\alpha \to 0+$. In fact, if the function $f_1(d,\alpha)$ is not continuous at $\alpha = 0$ what has to be obeyed is not $f_1(d,0) = 0$ but rather $\lim_{\alpha \to 0+} f_1(d,\alpha) = 0$. As shown in appendix B $\lim_{\alpha \to 0+} J_1(d,\alpha) = J_1(d,0)$. Therefore, the BC equation does not have a small $\alpha$ solution for large d. Indeed, also $\lim_{\alpha \to 0+} I_1(d,\alpha) \neq I_1(d,0)$ but still $\lim_{\alpha \to 0+} f_1(d,\alpha)$ does not vanish for the SE case.

Next, we tried to obtain a solution for $\alpha$ that is no necessarily small. This implies equating $f_1(d,\alpha)$ to zero for the SE and BC cases. Since $f_1(d,\alpha)$ is obtained explicitly this is straightforward and no solution was found.

The discussion above suggests two reasons for the appearance of severe numerical discrepancies. Taking into account the nature of the numerical difficulties we have recalculated the upper critical dimension and found from the SE equation $d_c = 3.7$ and from the BC equation $d_c = 4.3$ (a result which does not obey the Lassig inequality $d_c \leq 4$). It is interesting to note that the additional schemes by Blum and McKane [10] involve integrals that are more benign than $I_1$ or $J_1$, that are the cause of trouble in the SE and BC cases respectively. Therefore, it is quite natural that we recover their corresponding result with no deviations, for the three schemes in which they find an upper critical dimension (schemes 2, 3, 4 - eqs. (11)-(13) above).

Another important motivation for this work was the growing interest in mode-coupling and self-consistent-expansion approaches in many fields of non-linear science. The success of these approaches to handle problems that are not otherwise manageable is of course the important reason for their popularity. In this paper we intended to make a contribution for the development of these approaches by laying more rigorous foundations for them. We showed that in the KPZ problem the lack of such a rigorous understanding led to the publication of contradicting results for the values of the upper critical dimension as well as for the critical exponents (in high dimensions) by researchers using exactly the same equations. Such a situation indicates the need of a better understanding of the equations derived from these approaches, an understanding we offer. We believe that once this issue is settled a more consistent, fruitful and rapid development of the mode-coupling and self-consistent approaches will be possible.

We thank Sam Edwards for interesting discussions, and Mike Cates for his helpful comments, and for drawing our attention to the existence of the erratum [15] that yields a higher UCD.

REFERENCES

[1]   M. Kardar, G. Parisi and Y.-C. Zhang, Phys. Rev. Lett. **56**, 889 (1986).

[2]   S. F. Edwards and D. R. Wilkinson , Proc. R. Soc. London Ser. A **381**, 17 (1982).

[3]   T. Ala-Nissila, Phys. Rev. Lett. **80**, 887 (1998).

J. M. Kim, *ibid.*, **80**, 888 (1998).

M. Lassig and H. Kinzelbach, *ibid.*, **80**, 889 (1998).

[4]   T. Ala-Nissila, T. Hjelt, J. M. Kosterlitz and O. Venoloinen, J. Stat. Phys. **72**, 207 (1993).

L. H. Tang *et al.*, Phys. Rev. A **45**, 7162 (1992).

[5]   E. Perlsman and M. Schwartz, Physica A **234**, 523 (1996).

[6]   C. Castellano, M. Marsili and L. Pietronero, Phys. Rev. Lett. **80**, 3527 (1998).

C. Castellano, A. Gabrielli, M. Marsili, M. A. Munoz and L. Pietronero, Phys. Rev. E **58**, 5209 (1998).

[7]   T. Halpin-Healy, Phys. Rev. A **42**, 711 (1990).

[8]   M. Schwartz and S. F. Edwards, Europhys. Lett. **20**, 301 (1992).

[9]   M. Schwartz and S. F. Edwards, Phys. Rev. E **57**, 5730 (1998).

[10]   T. Blum and A. J. McKane, Phys. Rev. E **52**, 4741 (1995).

[11]   J-P. Bouchaud and M. E. Cates, Phys. Rev. E **47**, 1455 (1993).

[12]   J. P. Doherty *et al.*, Phys. Rev. Lett. **72**, 2041 (1994).

M. A. Moore *et al.*, Phys. Rev. Lett. **74**, 4257 (1995).

[13]   M. Lassig, Nucl. Phys. B **448**, 559 (1995).

M. Lassig and H. Kinzelbach, Phys. Rev. Lett. **78**, 903 (1997).

[14]   J. Cook and B. Derrida, J. Phys. A **23**, 1523 (1990).

[15]  J-P. Bouchaud and M. E. Cates (erratum), Phys. Rev. E **48**, 653 (E) (1993).

Appendix A

In this appendix we derive the large d asymptotic expansion for the integrals $I_1(d,\alpha)$, $I_2(d,\alpha)$, $J_1(d,\alpha)$ and $J_2(d,\alpha)$. We begin with $I_1(d,\alpha)$ and $I_2(d,\alpha)$. First, it can easily shown that the d-dimensional integration in $I_1(d,\alpha)$ (eq. (6)) can be reduced to the following 2-dimensional integral (using hyperspherical coordinates)

$$I_1(d,\alpha) = \int d^d t \frac{[\vec{t}\cdot(\hat{e}-\vec{t})][\vec{t}\cdot\hat{e}]t^{-(d+2\alpha)}}{t^{2-\alpha} + |\hat{e}-\vec{t}|^{1-\alpha/2} + 1} =$$

$$= \Omega(d-1)\int_0^\infty t^{d-1}dt \int_0^\pi (\sin\theta)^{d-2} d\theta \frac{[t\cos\theta - t^2][t\cos\theta]t^{-(d+2\alpha)}}{t^{2-\alpha} + |t^2 - 2t\cos\theta + 1|^{1-\alpha/2} + 1} \quad \text{(A1)}$$

Where $\Omega(d-1)$ is the surface area of the unit sphere in (d-1)-dimensions.

After substituting $x = \cos\theta$ we get

$$I_1(d,\alpha) = \Omega(d-1)\int_0^\infty dt \int_{-1}^1 dx (1-x^2)^{\frac{d-3}{2}} \frac{(x-t)xt^{1-2\alpha}}{t^{2-\alpha} + |t^2 - 2tx + 1|^{1-\alpha/2} + 1} \quad \text{(A2)}$$

And similarly for $I_2$

$$I_2(d,\alpha) = \Omega(d-1)\int_0^\infty dt \int_{-1}^1 dx (1-x^2)^{\frac{d-3}{2}} \frac{(x-t)^2 t^{1-2\alpha}(t^2 - 2tx + 1)^{-(d/2+\alpha)}}{t^{2-\alpha} + |t^2 - 2tx + 1|^{1-\alpha/2} + 1}. \quad \text{(A3)}$$

To evaluate the integrals for large d we expand the d-dependent parts of the integrands and perform the x-integration (Laplace method). This gives to a leading order in d

$$I_1(d,\alpha) = \Omega(d-1)\frac{\sqrt{2\pi}}{d^{3/2}} \int_0^\infty dt \frac{t^{1-2\alpha}}{t^{2-\alpha} + (t^2+1)^{1-\alpha/2} + 1} \times$$

$$\times \left[1 - (2-\alpha)\frac{t^2}{t^2+1} \frac{(t^2+1)^{1-\alpha/2}}{t^{2-\alpha} + (t^2+1)^{1-\alpha/2} + 1}\right] \equiv \Omega(d-1)\frac{\sqrt{2\pi}}{d^{3/2}} K_1(\alpha) \quad (A4)$$

To evaluate $I_2$, we note that to apply the Laplace method we have to find the maximum of the product $(1-x^2)(t^2 - 2tx + 1)^{-1}$ as a function of x in the range $-1 \leq x \leq 1$ and for fixed t. It turns out that for $0 \leq t \leq 1$ the maximum is attained at $x = t$ and for $1 \leq t < \infty$ the maximum is attained at $x = 1/t$. Therefore, we split $\int_0^\infty dt$ into two parts $\int_0^1 dt$ and $\int_1^\infty dt$, where for each of these t regions the x integral is performed separately and $I_2$ is given by

$$I_2(d,\alpha) = \Omega(d-1)\frac{\sqrt{2\pi}}{d^{3/2}} \int_0^1 dt \frac{t^{1-2\alpha}(1-t^2)^{-\alpha}}{t^{2-\alpha} + (1-t^2)^{1-\alpha/2} + 1} +$$

$$+ \Omega(d-1)\frac{\sqrt{2\pi}}{d^{1/2}} \int_1^\infty dt \frac{t^{1-2\alpha-d}(t^2-1)^{1-\alpha}}{t^{2-\alpha} + (1-t^2)^{1-\alpha/2} + 1} \quad (A5)$$

Notice that in the second contribution to $I_2$, the integral over t still has a d dependence. So next we calculate this contribution for large d

$$I_2(d,\alpha) = \Omega(d-1)\frac{\sqrt{2\pi}}{d^{3/2}} \int_0^1 dt \frac{t^{1-2\alpha}(1-t^2)^{-\alpha}}{t^{2-\alpha} + (1-t^2)^{1-\alpha/2} + 1} +$$

$$+ \Omega(d-1)\frac{\sqrt{2\pi}}{d^{1/2}} 2^{-\alpha} B(2-\alpha, d-(2-\alpha)) \quad (A6)$$

where $B(x, y)$ is the beta function.

Now, $B(2-\alpha, d-(2-\alpha)) = \Gamma(2-\alpha) \dfrac{\Gamma(d-(2-\alpha))}{\Gamma(d)}$, therefore for large d we use

the stirling formula to obtain the following d dependence of $I_2$

$$I_2(d, \alpha) = \Omega(d-1) \frac{\sqrt{2\pi}}{d^{3/2}} \int_0^1 dt \, \frac{t^{1-2\alpha}(1-t^2)^{-\alpha}}{t^{2-\alpha} + (1-t^2)^{1-\alpha/2} + 1} + \\ + \Omega(d-1) \frac{\sqrt{2\pi}}{d^{5/2-\alpha}} 2^{-\alpha} \Gamma(2-\alpha)$$

(A7)

Since $\alpha < 1$, the leading order of $I_2(d, \alpha)$ is given by

$$I_2(d, \alpha) = \Omega(d-1) \frac{\sqrt{2\pi}}{d^{3/2}} K_2(\alpha),$$

where

$$K_2(\alpha) = \int_0^1 dt \, \frac{t^{1-2\alpha}(1-t^2)^{-\alpha}}{t^{2-\alpha} + (1-t^2)^{1-\alpha/2} + 1}. \quad (A8)$$

The coefficients $K_1(\alpha)$ and $K_2(\alpha)$ can be easily evaluated at $\alpha = 0$ to yield

$$I_1(d, \alpha) = \Omega(d-1) \frac{\sqrt{2\pi}}{4d^{3/2}} \quad (A9)$$

$$I_2(d, \alpha) = \Omega(d-1) \frac{\sqrt{2\pi}}{4d^{3/2}}. \quad (A10)$$

Applying exactly the same procedure to $J_1$ and $J_2$ yields to leading order in 1/d,

$$J_1(d,\alpha) = \Omega(d-1)\frac{\sqrt{2\pi}}{d^{3/2}}L_1(\alpha), \qquad (A11)$$

where $L_1(\alpha)$ is obtained from $K_1(\alpha)$ by replacing the expression

$\left[t^{2-\alpha} + (t^2+1)^{1-\alpha/2} + 1\right]$ in (A4) by $\left[t^{2-\alpha} + (t^2+1)^{1-\alpha/2}\right]$, and

$$J_2(d,\alpha) = \Omega(d-1)\frac{\sqrt{2\pi}}{d^{3/2}}L_2(\alpha), \qquad (A12)$$

where again $L_2$ is obtained from $K_2$ by the same replacement.

The coefficients $L_1(\alpha)$ and $L_2(\alpha)$ can now be evaluated at $\alpha = 0$ to yield

$$J_1(d,\alpha) = \Omega(d-1)\frac{\sqrt{2\pi}}{4d^{3/2}}, \qquad (A13)$$

and

$$J_2(d,\alpha) = \Omega(d-1)\frac{\sqrt{2\pi}}{2d^{3/2}}. \qquad (A14)$$

Appendix B

In this appendix we show that

$$\lim_{\alpha \to 0+} J_1(d,\alpha) \neq J_1(d,0). \qquad (B1)$$

We begin by considering $L_1(\alpha)$ defined in (A11)

$$L_1(\alpha) = \int_0^\infty dt \frac{t^{1-2\alpha}}{t^{2-\alpha} + (t^2+1)^{1-\alpha/2}} \left[ 1 - \left(1 - \frac{\alpha}{2}\right) \frac{t^2}{t^2+1} \frac{2(t^2+1)^{1-\alpha/2}}{t^{2-\alpha} + (t^2+1)^{1-\alpha/2}} \right] \qquad (B2)$$

and we define $j(t,\alpha)$ to be the integrand in this integral. Now, we want to calculate

$$\lim_{\alpha \to 0+} L_1(\alpha) = \lim_{\alpha \to 0+} \int_0^\infty j(t,\alpha) dt. \qquad (B3)$$

Problems in evaluating the limit can arise from the tail of the integral. Therefore, we break the integral into two parts

$$\int_0^\infty j(t,\alpha) dt = \int_0^R j(t,\alpha) dt + \int_R^\infty j(t,\alpha) dt \qquad (B4)$$

where R can be arbitrarily large, and calculate each part separately.

the first part, i.e. $\int_0^R j(t,\alpha) dt$, we can take the $\lim_{\alpha \to 0+}$ into the integral because the interval, over which the integration is performed, is finite.

$$\lim_{\alpha \to 0+} \int_0^R j(t,\alpha)\,dt = \int_0^R j(t,0)\,dt = \int_0^R \frac{t}{2t^2+1}\left[1 - \frac{t^2}{2t^2+1}\frac{2(t^2+1)}{2t^2+1}\right]dt = \frac{1}{4}\frac{2R^2}{2R^2+1} \qquad (B5)$$

In the second contribution to the integral in eq. (B4) we expand the integrand for large t's, assuming we chose a large enough R. We obtain

$$\begin{aligned}j(t,\alpha) &= \frac{t^{-1-\alpha}}{2}\frac{2t^{2-\alpha}}{t^{2-\alpha}+(t^2+1)^{1-\alpha/2}}\left[1-\left(1-\frac{\alpha}{2}\right)\frac{t^2}{t^2+1}\frac{2(t^2+1)^{1-\alpha/2}}{t^{2-\alpha}+(t^2+1)^{1-\alpha/2}}\right] = \\ &= \frac{t^{-1-\alpha}}{2}\left[\frac{\alpha}{2} + \frac{1}{4t^2}(2\alpha-1)(\alpha-2) + O\!\left(\frac{1}{t^4}\right)\right]\end{aligned} \qquad (B6)$$

Performing the t-integration we get

$$\int_R^\infty j(t,\alpha)\,dt = \frac{1}{4}R^{-\alpha} + \frac{(2\alpha-1)(\alpha-2)}{8(\alpha+2)}R^{-2-\alpha} + O\!\left(\frac{1}{R^{4+\alpha}}\right) \qquad (B7)$$

Now, we can take easily take the $\lim_{\alpha \to 0+}$

$$\int_R^\infty j(t,\alpha)\,dt \xrightarrow[\alpha \to 0+]{} \frac{1}{4} + \frac{1}{8R^2} + O\!\left(\frac{1}{R^4}\right). \qquad (B8)$$

At this point we remember that R can be taken to be arbitrarily large, so we obtain

$$\lim_{\alpha \to 0+} \int_0^\infty j(t,\alpha)\,dt = \frac{1}{4}\frac{2R^2}{2R^2+1} + \frac{1}{4} + \frac{1}{8R^2} \xrightarrow[R\to\infty]{} \frac{1}{4} + \frac{1}{4} = \frac{1}{2} \qquad (B9)$$

and therefore $\lim_{\alpha \to 0+} L_1(\alpha) \neq L_1(0)$. More specifically

$$\lim_{\alpha \to 0+} L_1(\alpha) = 2L_1(0). \qquad (B10)$$